%% file: main.tex
\documentclass[5p]{./JWS/elsarticle}

\makeatletter
\def\ps@pprintTitle{%
 \let\@oddhead\@empty
 \let\@evenhead\@empty
 \def\@oddfoot{\textit{Accepted manuscript, Journal of Web Semantics 2020, 
DOI: \url{https://doi.org/10.1016/j.websem.2020.100586}}
}%
 \let\@evenfoot\@oddfoot}
\makeatother

\usepackage{lineno,hyperref}
\usepackage{todonotes}
\presetkeys{todonotes}{inline}{}
\usepackage{booktabs}
\usepackage{amsmath}
\usepackage{amsfonts}
\usepackage{subcaption} 
\usepackage{soul}

\usepackage{flushend}
\usepackage{multicol}

\usepackage{tabularx}
\usepackage{graphicx}
\usepackage{adjustbox}
\usepackage{balance}

\journal{Journal of Web Semantics}

\usepackage[normalem]{ulem}

\begin{document}

\begin{frontmatter}

\title{IQA: Interactive Query Construction in Semantic Question Answering Systems\tnoteref{t1}}

\tnotetext[t1]{
\copyright~2020. This manuscript version is made available under the CC-BY-NC-ND 4.0 license \url{http://creativecommons.org/licenses/by-nc-nd/4.0/}
}

\author[A]{Hamid Zafar}
\ead{hzafarta@cs.uni-bonn.de}

\author[A,B]{Mohnish Dubey}
\ead{dubey@cs.uni-bonn.de}

\author[A,B]{Jens Lehmann}
\ead{jens.lehmann@cs.uni-bonn.de}

\author[C]{Elena Demidova\corref{correspondingauthor}}
\ead{demidova@L3S.de}

\cortext[correspondingauthor]{Corresponding author}

\address[A]{Smart Data Analytics Group (SDA), University of Bonn, Germany}
\address[B]{Enterprise Information Systems Department, Fraunhofer IAIS, Germany
 }
\address[C]{L3S Research Center, Leibniz Universit\"at Hannover, Germany}


\begin{abstract}
Semantic Question Answering (SQA) systems automatically interpret user questions expressed in a natural language in terms of semantic queries. This process involves uncertainty, such that the resulting queries do not always accurately match the user intent, especially for more complex and less common questions. 
In this article, we aim to empower users in guiding SQA systems towards the intended semantic queries through interaction. We introduce IQA - an interaction scheme for SQA pipelines. This scheme facilitates seamless integration of user feedback in the question answering process and relies on Option Gain - a novel metric that enables efficient and intuitive user interaction. Our evaluation shows that using the proposed scheme, even a small number of user interactions can lead to significant improvements in the performance of SQA systems.
\end{abstract}


\end{frontmatter}


\input{01_introduction.tex}

\input{02_formalization.tex}

\input{03_interaction_scheme.tex}

\input{04_realization.tex}

\input{05a_evaluation_setup.tex}

\input{05b_evaluation_results.tex}

 \input{06_relatedwork.tex}

 \input{07_conclusion.tex}

\bibliographystyle{elsarticle-num}

\bibliography{main}

\end{document}

%% file: 01_introduction.tex
\section{Introduction}
\label{sec:introduction}

Openly available large-scale knowledge graphs such as DBpedia \cite{LehmannIJJKMHMK15}, Wikidata \cite{VrandecicK14}, 
YAGO \cite{HoffartSBW13} and EventKG \cite{GottschalkD18}, \cite{gottschalk2019eventkg} have evolved as the key reference sources of information and knowledge regarding real-world entities, events, and facts on the Web. The flexibility of the RDF-based knowledge representation, the large-scale editor base of popular knowledge graphs, and recent advances in the automatic knowledge graph completion methods lead to a growth of the data and the schema layers of these graphs at an unprecedented scale, with schemas including thousands of types and relations \cite{Paulheim17}. As a result, the information contained in the knowledge graphs is very hard to query, in particular, due to the large scale, the heterogeneity of the entities, and the variety of their schema descriptions.

Semantic Question Answering (SQA) is the key technology to facilitate end-users to query knowledge graphs using natural language interfaces. In recent years, a large number of SQA approaches have been developed \cite{HoffnerWMULN17}. The objective of these approaches is to automatically interpret a user question formulated in a natural language as a semantic query (typically expressed in the SPARQL query language), which is then executed against the knowledge graph to obtain the results. Current SQA approaches are capable of effectively answering rather simple factual questions that contain a limited number of entities and relations.

In the case of complex questions, i.e., questions that involve multiple entities and relations, the performance of the existing SQA approaches is still limited. These limitations can, to a large extent, be attributed to the inherent uncertainty associated with the results of the individual pipeline components along with the propagation of errors of the component results through the entire SQA pipeline. 
This uncertainty often leads to imprecise question interpretations, especially for complex questions.

\begin{figure*}
  \centering
   \includegraphics[width=0.85\textwidth]{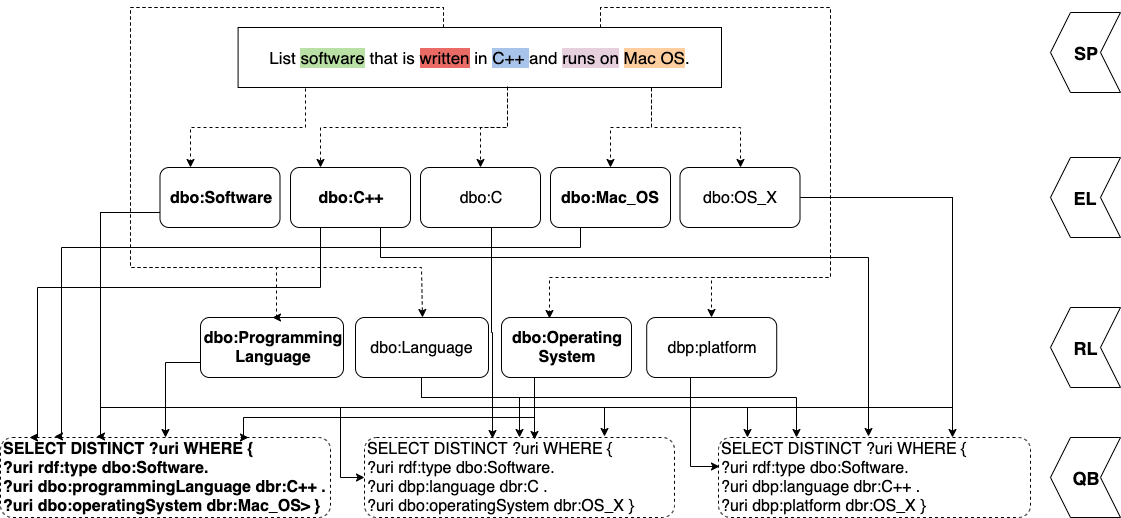}
    \caption{An example transformation of a question from the LC-QuAD dataset in possible semantic queries over the DBpedia knowledge graph using an SQA pipeline consisting of a Shallow Parser (SP), an Entity Linker (EL), a Relation Linker (RL) and a Query Builder (QB).} 
\label{fig:example}
\end{figure*}

Figure~\ref{fig:example} illustrates this problem using an example question from 
LC-QuAD \cite{LC-QuAD} - a state-of-the-art dataset for evaluation of Semantic Question Answering systems: \textit{``List software that is written in C++ and runs on Mac OS.''}.
An SQA pipeline incrementally transforms the input question in a semantic query, using components such as a \textit{Shallow Parser} (SP), an \textit{Entity Linker} (EL), a \textit{Relation Linker} (RL) and a \textit{Query Builder} (QB). 
First, the Shallow Parser identifies keyword phrases ``software'', ``written'', ``C++'', ``runs'' and ``Mac OS''.
Then the Entity Linker and Relation Linker 
map these keyword phrases to the entities and relations in the DBpedia knowledge graph.
To obtain correct interpretation, the Entity Linker should link the keyword phrase ``C++'' to the entity \textit{dbr:C++}\footnote{\url{http://dbpedia.org/resource/C++}}, the programming language and ``Mac OS'' to the entity \textit{dbr:Mac\_OS}\footnote{\url{http://dbpedia.org/resource/Mac\_OS}}, the operating system.
The Entity Linker should not confuse ``C++'' with e.g., \textit{dbr:C}\footnote{\url{http://dbpedia.org/resource/C}}, another programming language.
The Relation Linker should link the keyword phrases ``written'' to the relation \\ \textit{dbo:programmingLanguage}\footnote{\url{http://dbpedia.org/ontology/programmingLanguage}} and ``runs on'' to the relation \textit{dbo:operatingSystem}\footnote{\url{http://dbpedia.org/ontology/operatingSystem}}. 
Here, the task of relation linking is particularly difficult due to the lexical gap, the required domain knowledge, and the ambiguity of the candidates. 
To reduce the number of candidates, the Relation Linker can rely on the Entity Linker results, e.g., by taking into account the relations of the linked entities in the knowledge graph. 
Finally, the Query Builder component utilizes the results of the Entity Linker and Relation Linker to build the semantic query. 
Errors in the results of the Entity Linker and Relation Linker can often lead to the misinterpretation of the user question. With an increasing number of entities and relations mentioned in the user question, the likelihood of such errors increases.

The objective of this article is to address the limitations of the existing SQA approaches in answering complex questions
through the provision of a novel user interaction scheme. 
While other domains like Information Retrieval and keyword search over structured data take significant advantage of user interaction models (e.g., \cite{Demidova:2013}), such models are not yet widely adopted in the context of Semantic Question Answering. The proposed IQA scheme can be particularly beneficial in answering complex questions when the intended semantic interpretation of the question cannot be accurately inferred using automatic methods. From the algorithmic perspective, this scheme can facilitate SQA systems to reduce uncertainty during the query interpretation process efficiently. 
From the user perspective, this scheme can empower users in effectively guiding SQA algorithms towards the intended results.

Given an SQA pipeline and a user question, the goal of IQA is to facilitate an efficient and intuitive generation of the intended question interpretation through user interaction.
The proposed interaction scheme incrementally refines user questions in the intended semantic queries by requesting user feedback on several items called \textit{interaction options}.
The main challenge to be addressed here is the trade-off between the
efficiency and the usability in the interaction scheme. 
In this context, efficiency refers to the minimization of the interaction cost (i.e., the number of requests for user feedback). The usability means the ease of use/understandability of the interaction options. 
To the best of our knowledge, none of the state-of-the-art SQA systems support user interaction in Semantic Question Answering in the way envisioned in this article.

Overall, in this article we make the following contributions: 
\begin{itemize}
    \item We provide a formalization of an SQA pipeline, which captures the dependency of the pipeline components, and facilitates generalization of the proposed interaction scheme to a wide range of SQA systems. 
    \item We present a probabilistic foundation to estimate the likelihood of the generated question interpretations and interaction options. This model builds a basis for the systematic generation of effective interaction options in a variety of categories. 
    \item We propose a user interaction scheme that seamlessly incorporates user feedback in the Semantic Question Answering process to reduce uncertainty efficiently. We adopt a cost-sensitive decision tree to balance the trade-off between usability and efficiency of the options in the interaction process. 
    \item We incorporate the usability of interaction options into a new metric, \textit{Option Gain}, that balances the usability and efficiency of interaction options and facilitates the selection of interaction options that are efficient and intuitive for the user. 
    \item We showcase an instantiation of the proposed user interaction scheme in a web-based IQA prototype while utilizing existing components developed by the SQA community. 
\end{itemize}

We demonstrate the effectiveness and efficiency of the proposed interaction scheme for Semantic Question Answering in an extensive experimental evaluation and a user study. 
Our evaluation results on LC-QuAD, an established dataset for the assessment of Semantic Question Answering systems, demonstrate that IQA can significantly improve the effectiveness, efficiency, and usability of Semantic Question Answering systems for complex questions.
In particular, the IQA-OG configuration that adopts Option Gain achieves an increase of up to 20 percentage points in terms of $F_1$ score compared to the baselines on a subset of LC-QuAD utilized in the user study. Furthermore, this configuration enhances the ease of use as reported by the users.

We organize the rest of the article as follows:
First, we formalize the concept of Semantic Question Answering pipeline in Section \ref{sec:formalization}. Then, in Section \ref{sec:ui-scheme} we present the user interaction 
scheme of IQA. Following that, we describe the realization of the IQA pipeline in Section \ref{sec:realization}. 
The evaluation setup is described in Section \ref{sec:evaluation-setup}.
Our evaluation results are presented in Section \ref{sec:evaluation_results}.
Section \ref{sec:relatedwork} discusses related work. We provide a conclusion in Section \ref{sec:conclusion}.

%% file: 02_formalization.tex
\section{Formalization of an SQA Pipeline}
\label{sec:formalization}

A Semantic Question Answering pipeline (denoted as ``SQA pipeline\text{''} in the following) transforms a user question specified in a natural language into a semantic query for the target knowledge graph. In this section, we present a formalization of an SQA pipeline that abstracts from the particular implementation.
Notations frequently used in the article are summarized in Table \ref{tab:notation}.

\begin{table}[t]
\centering
\caption{Summary of frequently used notations.}
\label{tab:notation}
\begin{adjustbox}{width=0.45\textwidth}
\begin{tabular}{@{}ll@{}}
\toprule
\textbf{Notation} & \textbf{Description} \\ \midrule
$Q= (q_{NL}, QN)$ & a representation of the user question \\
$q_{NL}$  & a user question as a natural language expression\\
 $QN$ & a multiset of information nuggets\\
 $QI$ & a partial question interpretation \\ 
 $CQI$ & a complete question interpretation \\ 
 $plc$ & an interpretation function \\
 $QIS$ & the question interpretation space \\
  $IO$ & an interaction option \\
   $OG$ & Option Gain \\
 $IG$ & Information Gain\\
\bottomrule
\end{tabular}
\end{adjustbox}
\end{table}

\subsection{Basic Concepts}

The goal of Semantic Question Answering is to transform a user question expressed in a natural language into a semantic query for the target knowledge graph. 
In the following, we formalize the concepts of the knowledge graph, the user question, and the semantic query. 

A \emph{knowledge graph} $\mathcal{KG} = (V, L, E, T)$ consists of a set $V$ of entities, a set $L$ of literals, 
a set $E$ of properties and a set $T \subseteq V \times E \times (V \cup L)$ of triples.

The entities in $V$ represent real-world entities and concepts. 
The properties in $E$ represent relations connecting two entities or an entity and a literal value.

A \textit{user question} $Q= (q_{NL},QN)$ is a tuple that represents user input. $q_{NL}$ is the initial user question expressed in a natural language. 
$QN =\{n_{1}, \ldots, n_{m}\}$ is a multiset of information nuggets mentioned in the user question. 

Information nuggets can include surface forms of named entities, concepts, and relations mentioned in $q_{NL}$. 
Information nuggets can be extracted from $q_{NL}$ using information extraction techniques such as shallow parsing.

For example, consider the question:
\begin{flalign*}
&q_{NL}=\\
&\text{\textit{``List software that is written in C++ and}}\\ 
&\text{\textit{runs on Mac OS.\text{''}}}
\end{flalign*}
This question can be transformed into the following set of information nuggets: 
\begin{flalign*}
&QN = \{\\
&``software\text{''},~``written\text{''},~``C++\text{''},~``runs\text{''},~``Mac~OS\text{''}\}.
\end{flalign*}

In the process of Semantic Question Answering, information nuggets mentioned in the user question are interpreted as elements of the knowledge graph. 
A \textit{nugget interpretation} $ni$ is a mapping from an information nugget $n \in QN$ to an element of the knowledge graph $\mathcal{KG}$. 
An information nugget can be interpreted as an entity, a literal, a property, a single triple, or a set of triples.

For example, the nugget interpretation:
\begin{flalign*}
&ni_0=\{``software\text{''} \mapsto dbo:Software\}
\end{flalign*}
maps the information nugget 
\textit{``software\text{''}} to the entity \\ \textit{``dbo:Software\text{''}} of the knowledge graph.
Other examples of nugget interpretations include: 
\begin{flalign*}
&ni_1=\{``written\text{''} \mapsto dbo:programmingLanguage\}, \\
&ni_2=\{``C++\text{''} \mapsto dbr:C\text{++}\}, \\
&ni_3=\{``runs\text{''} \mapsto dbo:operatingSystem\}, \\
&ni_4=\{``Mac ~ OS\text{''} \mapsto dbr:Mac\_OS\}.
\end{flalign*}

When an SQA pipeline transforms the user question $Q$ into a semantic query, the pipeline components can generate intermediate interpretation results that include several nugget interpretations. We refer to such intermediate results as \textit{partial question interpretations}. 
More formally:

A \textit{partial question interpretation} 
$QI=\{ni_{1}, \ldots, ni_{r}\}$ is a set of nugget interpretations that interpret a (sub)set of the information nuggets 
contained in $QN$.

For example, a partial question interpretation  
\begin{flalign*}
&QI=\{ \\
& \quad \{``C\text{++}\text{''} \mapsto dbr:C\text{++}\}, \\
& \quad \{``Mac ~ OS\text{''} \mapsto dbr:Mac\_OS\}
\}
\end{flalign*}
includes specific interpretations of two information nuggets representing entity surface forms in the user question. 

Partial question interpretations serve as a basis for building semantic queries.

A semantic query $CQI$ is a \textit{complete question interpretation} that represents the user question as a whole. 
Intuitively, a $CQI$ includes the elements of the knowledge graph that correspond to the nugget interpretations 
and connects them in a graph pattern. 

Formally, a \emph{complete question interpretation} 
$CQI=(QI, AT, QG)$ is a tuple that consists of a set of nugget interpretations $QI$, 
an answer type $AT$ and a query graph $QG$.
The answer type $AT$ 
is an element of \{``ASK\text{''}, ``SELECT\text{''}, ``COUNT\text{''}\}.
Given a knowledge graph $\mathcal{KG} = (V, L, E, T)$, a query graph $QG=(V', L', E', U, T')$ is a graph pattern 
such that: 
$V' \subset V$ is a set of entities, 
$L' \subset L$ is a set of literals,  
$E' \subset E$ is a set of properties, 
$U$ is a set of variables and 
$T' \subset (V' \cup U) \times (E' \cup U) \times (V' \cup L' \cup  U) $ is a set of triple patterns. 

For example, $CQI_1 = (QI_1, QG_1, AT_1)$, is a complete question interpretation of the example question presented above, where:
$AT_1=``SELECT\text{''}$, 

\begin{flalign*}
&QI_1=\{ \\
& \quad\{``software\text{''} \mapsto dbo:Software\}, \\
&\quad\{``written\text{''} \mapsto  dbo:programmingLanguage\}, \\
&\quad\{``C++\text{''} \mapsto dbr:C\text{++}\}, \\
&\quad\{``runs\text{''} \mapsto dbo:operatingSystem\}, \\
&\quad\{``Mac ~ OS\text{''} \mapsto dbr:Mac\_OS\}
\},
\end{flalign*}
and 
\begin{flalign*}
&QG_1= ( \\
&V'=\{dbo:Software, dbr:C\text{++}, dbr:Mac\_OS\}, \\
&L'= \emptyset, \\
&E'= \{rdf:type, dbo:programmingLanguage,\\ &dbo:operatingSystem\}, \\
&U= \{?uri\}, \\
%
&T'= \{ \\
&\quad?uri~rdf:type~dbo:Software, \\
&\quad?uri~dbo:programmingLanguage~dbr:C\text{++}, \\
&\quad?uri~dbo:operatingSystem~dbr:Mac\_OS. 
\}).
\end{flalign*}

To retrieve answers from a knowledge graph, a complete question interpretation can be translated into a query in the SPARQL query language\footnote{\url{https://www.w3.org/TR/sparql11-query}}.  
For example, the following SPARQL query corresponds to the complete question interpretation 
of the example question presented above: 

\begin{flalign*}
&SELECT~?uri~WHERE~\{ \\
&?uri~rdf:type~dbo:Software. \\
&?uri~dbo:programmingLanguage~dbr:C\text{++}. \\
&?uri~dbo:operatingSystem~dbr:Mac\_OS. \}
\end{flalign*}

Note that a complete question interpretation does not necessarily include interpretations of all information nuggets extracted from the user question. This is because information nuggets in $QN$ can potentially contain redundant information. 

\subsection{Semantic Question Answering Pipeline}

A typical Semantic Question Answering pipeline considered in this article consists of: 
1.) a shallow parser $plc_{sp}$ constructing information nuggets, 
2.) linkers $plc_{link}$: here we support different options of entity, relation and class linking separately or jointly -- so there can be one or multiple linkers, and
3.) a query builder $plc_{qb}$ creating complete question interpretations.

More formally, a \textit{Semantic Question Answering pipeline} $PL$ is a list of components, where each component $plc \in PL$ implements an \textit{interpretation function}. 
The aim of an interpretation function is to incrementally transform the user question into candidate question interpretations.

 \begin{flalign}
 PL= 
 \Bigg(
 \Big( plc_{sp}\Big),
 \Big(
 ( plc_{link_1}), \ldots, ( plc_{link_p})\Big), 
 \Big(plc_{qb}\Big)
 \Bigg). 
 \label{eq:pipeline}
 \end{flalign}

A pipeline component $plc$ can generate multiple candidate interpretations. 

The component $plc_{sp}$ is a specific shallow parsing component at the first step of the pipeline, which transforms the user question 
into a set of information nuggets:
$plc_{sp} : \mathbb{Q_{NL}} \mapsto \mathbb{QN}$, 
where $\mathbb{Q_{NL}}$ is the set of natural language questions, and $\mathbb{QN}$ is the set of information nuggets.

A $plc_{link}$ component takes the user question and, optionally, an interpretation produced by the previous pipeline component as an input and produces a set of partial interpretations as an output: 
$plc_{link} : \mathbb{Q} \times \mathbb{QI} \mapsto \mathcal{P}(\mathbb{QI})$, 
where 
$\mathbb{Q}$ is the set of questions, $\mathbb{QI}$ is the set of partial question interpretations, and $\mathcal{P}$ is the power set constructor.
Examples of interpretation functions of the components $plc_{link}$
include entity linking, relation linking, and class linking.
There can be a single joint linking step or multiple individual linking steps. 
By supporting all of those scenarios, the interaction framework described in this article can be applied to a broader range of existing SQA frameworks. 

The component $plc_{qb}$ is a specific query building component at the last step of the pipeline, which transforms a partial question interpretation $QI$ into one or more complete question interpretations, i.e. 
$plc_{qb} : \mathbb{QI} \mapsto \mathbb{CQI}$, 
where $\mathbb{CQI}$ is the set of complete question interpretations.
Each question interpretation $QI \in \mathbb{QI}$ and 
$CQI \in \mathbb{CQI}$
is associated with a confidence score generated by the corresponding pipeline component.

Conceptually, as an SQA pipeline processes the user question, it incrementally generates a hierarchy of question interpretations, where partial question interpretations are the intermediate nodes, 
and complete question interpretations are the leaf nodes.

%% file: 03_interaction_scheme.tex
\section{IQA User Interaction Scheme}
\label{sec:ui-scheme}

Given a user question $Q$ and a large-scale knowledge graph $\mathcal{KG}$, 
a Semantic Question Answering pipeline $PL$ can generate 
a large number of possible complete question interpretations. 
We denote the set of all complete question interpretations of $Q$ 
generated by $PL$ given $\mathcal{KG}$ 
as a \textit{question interpretation space} $QIS$.

IQA facilitates an efficient and intuitive generation of the intended question interpretation through a user interaction scheme. 
In IQA, an \textit{interaction option} $IO$ is a unit adapted for user interaction.
The goal of the interaction scheme is to reduce the question interpretation space $QIS$ with each user interaction efficiently while providing intuitive interaction options.
Conceptually, the IQA interaction scheme resembles the induction of 
a cost-sensitive decision tree~\cite{lomax2013survey}, where 
the cost reflects the complexity and usability of the interaction options from the user perspective. 
We rely on the notion of Option Gain introduced later in this section to facilitate the usability and efficiency of the interaction scheme. 

\subsection{Interaction Options and Subsumption Relation}
\label{sec:interaction-options}

An \textit{interaction option} $IO$ is a unit adapted for user interaction
to reduce the question interpretation space $QIS$.
In IQA we group interaction options in the following categories: 1) nugget interpretations, 
2) superclasses and types of entities, 3) answer types of semantic queries, 
and 4) complete question interpretations (i.e., semantic queries). 

To facilitate an effective reduction of the question interpretation space $QIS$ by interaction, we establish a subsumption relation between interaction options and complete question interpretations.

We say that an interaction option $IO$ subsumes a complete question 
interpretation $CQI=(QI, AT, QG)$ if one of the following conditions applies: 

\begin{itemize}
\item[C1.] Interaction option $IO$ represents 
a nugget interpretation 
leading to the generation of the semantic query, namely: $IO \in QI$. 
\item[C2.] Interaction option $IO$ is a superclass or a type of an entity included in $CQI$: there must be a URI $x$ in the query graph $QG$ of the complete query interpretation $CQI$, for which a triple 
\begin{flalign*}
(x, rdfs:subClassOf, y),~ or & \\
(x, rdf:type, y)  &
\end{flalign*}
\noindent exists in the knowledge graph, and $y\equiv IO$. 
\item[C3.] Interaction option $IO$ represents the answer type of $CQI$: $IO \equiv AT$.
\item[C4.] Interaction option $IO$ is equivalent to the semantic query: $IO \equiv CQI$.
\end{itemize}

\subsection{Option Gain}
\label{sec:option-gain}

Interaction options vary concerning their complexity and usability. Complex interaction options can be difficult to understand for the users, potentially leading to an error-prone interaction process (i.e., wrong user decisions) and decreasing an overall user satisfaction. 

The key concept of the IQA interaction scheme is the \textit{Option Gain} $OG(IO)$.
Option Gain takes into account the $usability(IO)$ and the
efficiency of the interaction option $IO$ expressed using its Information Gain $IG(IO)$.
We define the Option Gain as:

\begin{align}
OG(IO) &=  usability(IO)^{\omega} \times IG(IO),  
\label{eq:OptionGain}
\end{align}

\noindent where $\omega \in \mathbb{N}$ is a parameter that controls the bias introduced by the usability of an interaction option IO in the interaction process, such that by $\omega=0$ the Option Gain corresponds to the Information Gain without the usability bias.

In IQA the usability of an interaction option is reflected through the usability score $usability(IO) \in [0,1]$, where $1$ corresponds to the most intuitive options and $0$ to the most complex options:

\begin{flalign}
usability(IO) &= \frac{1}{ 1+ complexity(IO)}.
\label{eq:usability}
\end{flalign}

The complexity of an interaction option $complexity(IO)$ can be characterized through the syntactic similarity of the interaction option to the initial user question, the degree of abstraction, and the structural complexity. 

Given the user question $Q$, the uncertainty of the question interpretation is the result of several factors, including: 
F1) the ambiguity of information nuggets in $Q$ and the resulting
uncertainty when interpreting these nuggets in a large-scale knowledge graph; 
F2) the uncertainty of the expected answer type; and
F3) a variety of possible graph structures connecting nugget interpretations in a semantic query.
Interaction options proposed in IQA aim to reduce this uncertainty.

In the following, we discuss the complexity estimation of the interaction options, which were introduced in Section~\ref{sec:interaction-options} above.

\begin{itemize}
\item[C1.] 
An interaction option $IO$ in this category is a nugget interpretation. 
Intuitively, an $IO$ syntactically similar to the nugget in the user question may appear familiar, and thus less complex, to the user. 
Therefore, we estimate the complexity of an option $IO$ in this category as the dissimilarity between the information nugget corresponding to the $IO$ in the user question and the representation (e.g., a label) of the $IO$ shown to the user in the interaction process.
We adopt the Longest Common Substring (LCS) as a string similarity metric, as this metric was shown to be suitable for short phrases~\cite{christen2006comparison}. 
\item[C2.] 
An interaction option in this category is a superclass or a type of an information nugget contained in the semantic query. 
The usability of such options depends on the degree of abstraction. We assume that less abstract categories such as ``person'' and ``actor'' can appear more intuitive to the users than more abstract categories, such as ``living thing''. 
To reflect this intuition, we measure the complexity of the interaction options in this category as the length of the shortest path between the $IO$ and the element of the knowledge graph that directly maps to the corresponding information nugget in the user question.
\item[C3.] 
An interaction option in this category represents an answer type of the semantic query. 
Given a relatively straightforward set of possible answer types, we set $complexity(IO)=0$ for the options in this category. 
\item[C4.] 
The interaction options in this category are semantic queries. 
Intuitively, more complex queries that include a high number of nugget interpretations can appear more difficult to understand from the user perspective. Therefore, we compute the complexity of an interaction option 
in this category as the number of nugget interpretations it includes.
\end{itemize}

\subsection{Information Gain}
\label{sec:information-gain}

For the computation of the Information Gain
of an interaction option in the question interpretation space  $QIS$,
we build upon the probabilistic model proposed in our previous work \cite{Demidova:2013}.
We summarize the computation of the Information Gain in the following.

Let $H(QIS)$ be the entropy of the probability distribution in the question interpretation space $QIS$.
The Information Gain of an interaction option $IG(IO)$ is computed as the entropy reduction given user feedback on $IO$.

Let $QIS_{IO}$ be the set of complete question interpretations in $QIS$ subsumed by $IO$, and $QIS_{\overline{IO}}$ be the set of all other complete question interpretations in $QIS$. 
Furthermore, let $P(IO)$ be the probability that the interaction option $IO$ subsumes the user-intended
complete question interpretation.

The entropy of the probability distribution in the question interpretation space $QIS$ is computed as:

\begin{align}
H (QIS)&= - \sum_{CQI \in QIS} \mkern-18mu P(CQI|Q, \mathcal{KG}) \times log_{2} P(CQI|Q, \mathcal{KG}).
\label{eq:Entropy}
\end{align}

Then, Information Gain of the interaction option is computed as
the uncertainty reduction provided by this option:

\begin{flalign}
IG(IO) &= H (QIS) -  \\ & \nonumber
 \Bigg( P(IO) \times H (QIS_{IO}) + P (\overline{IO}) \times H (QIS_{\overline{IO}})\Bigg).
\end{flalign}

The probability of an interaction option $P(IO)$ is computed as the sum of the probabilities of complete question interpretations subsumed by this option:

\begin{align}
P(IO) &= \sum_{CQI \in QIS_{IO}} P(CQI|Q, \mathcal{KG}).
\label{eq:ProbOption}
\end{align}

\subsection{Probability of Complete Question Interpretations}
\label{sec:probabilityCQI}

To estimate the probability $P(CQI|Q,\mathcal{KG})$ of the complete question interpretation 
$CQI=(QI, AT, OG)$ to be intended by the user, given the user question $Q=(q_{NL}, QN)$ and the knowledge graph $\mathcal{KG}$, we consider the following factors: 
1) the likelihood of the partial question interpretation $QI=\{ni_{1}, \ldots, ni_{r}\}$ 
from which $CQI$ was composed by the SQA pipeline, 
represented as $P(QI|Q,\mathcal{KG})$ and 
2) the probability of the graph structure $QG$ of the semantic query given the linguistic structure of the user question $q_{NL}$, 
represented as $P(QG|q_{NL},\mathcal{KG})$.

For mathematical simplification, similar to Na\"ive Bayes,
we assume that the probabilities of the nugget interpretations in the $QI$ from which $CQI$ is constructed, as well as the structure 
$QG$ of the resulting semantic query are mutually independent.
Although the resulting probability estimation is potentially not very precise, it leads 
to an adequate prediction of query relevance, as shown by our experiments. 

Then the probability $P(CQI|Q,\mathcal{KG})$ of the complete question interpretation $CQI$ can be estimated as:
%


\begin{flalign}
    P(CQI|Q,\mathcal{KG}) & \propto  \\  
    \left( \prod_{ni_{i} \in QI} P(ni_{i}|Q,\mathcal{KG})\right) \nonumber \times P(QG|q_{NL},\mathcal{KG}). 
    \label{eq:ProbInterpretation}
\end{flalign}

We estimate $P(ni_{i} |Q,\mathcal{KG})$ using the confidence score provided by the pipeline component that generates the nugget interpretation $ni_{i}$. 
$P(QG|q_{NL},\mathcal{KG})$ is estimated using the structural similarity between the graph structure of $CQI$ 
and the parse tree structure of the user question $q_{NL}$.
We provide more details regarding the computation later in Section \ref{sec:prob-estimation}.

\subsection{User Interaction Process}
\label{sec:interaction-process}

The conceptual process of the interactive question interpretation using a generic Semantic
Question Answering pipeline presented in Section \ref{sec:formalization} can be modeled as follows: 

\textbf{Step 1 (SQA Pipeline Execution):} The user issues the question $Q$.
The SQA pipeline is executed to generate 
the question interpretation space $QIS$. 

\textbf{Step 2 (Pre-Processing):} The 
partial and complete question interpretations generated by the pipeline are utilized to generate the interaction options. Then the subsumption relations between these options and the complete question interpretations in $QIS$ are established. 

\textbf{Step 3 (User Interaction):} 
At each step of the interaction process, the user is simultaneously presented with:
\begin{itemize}
\item
The interaction option $IO^{*}$ with the highest \textit{Option Gain}, and 
\item
The most likely complete question interpretation $CQI^{*}$ in the interpretation space $QIS$ (in a natural language and a semantic representation).
\end{itemize}

For simplicity, we model the interaction process as a list of binary user decisions, i.e., we assume that the user is presented with one interaction option at a time. 
In practice, this process can be generalized to present several interaction options simultaneously.

At each step of the interaction process, the user has the following means to interact with the system: 
\begin{itemize}
\item
Accept the interaction option $IO^{*}$, i.e., confirm that the presented interaction option correctly interprets (a part of) the question $Q$. 
\item
Reject the interaction option $IO^{*}$. 
\item
Accept the complete question interpretation $CQI^{*}$, i.e., confirm that this interpretation correctly reflects the intention of the question.
\end{itemize}

After each interaction, $CQI$s that do not comply with the user decision are removed from the question interpretation space $QIS$ using subsumption relation. 
The Option Gain of all the interaction options is recomputed. The interaction process continues with the currently top-scored $IO^{*}$ and $CQI^{*}$.

The interaction process for a question terminates if one of the following applies: 
\begin{itemize}
\item
The user accepts the complete question interpretation $CQI^{*}$. 
\item
The question interpretation space $QIS$ is empty, i.e., the correct interpretation cannot be identified given user feedback.
\item
The user terminates the process.
\item
The number of interactions or the time spent by the user reached a threshold. 
\end{itemize}

%% file: 04_realization.tex
\section{Realization}
\label{sec:realization}

In this section, we present the realization of the proposed IQA approach presented in Section \ref{sec:ui-scheme}, 
including in particular an IQA pipeline implementation and a prototypical user interface adopted in the user evaluation.
Note that our approach is independent of any specific implementation of the SQA pipeline
formalized in Section \ref{sec:formalization}. 

\subsection{IQA Pipeline}
\label{sec:iqa-pipeline}

\begin{figure*}[h!]
\includegraphics[width=\textwidth]{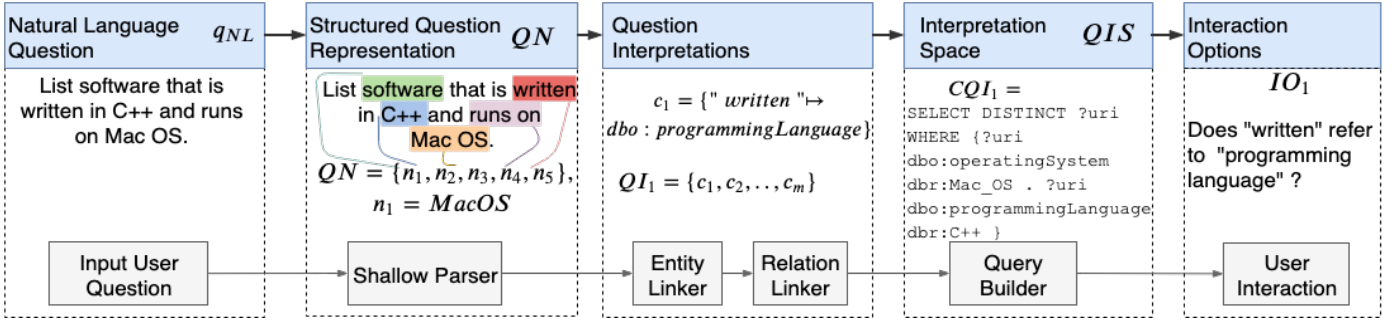}
\caption{An Interactive Question Answering (IQA) pipeline.} 
\label{fig:pipeline}
\end{figure*}

The Semantic Question Answering pipeline of IQA instantiated in this work is illustrated in Figure \ref{fig:pipeline}.
This pipeline consists of four components, namely a Shallow Parser, an Entity Linker, a Relation Linker, and a Query Builder.

With the IQA pipeline, we aim to generate several relevant candidate question interpretations 
to build the interpretation space $QIS$, to facilitate the user interaction scheme. 
This method is different from the state-of-the-art SQA approaches such as
``WDAqua'' \cite{DBLP:journals/corr/abs-1803-00832} 
aimed to generate only one, the most likely question interpretation. 

To increase the recall of relevant question interpretations generated by the IQA pipeline, 
we leverage multiple independent tools in each pipeline step to obtain complementary candidates. 
The output of each pipeline component is the union of all candidates produced by the individual tools. 
This approach increases the recall of the candidates generated in each pipeline component. 
Furthermore, it increases the overall recall of the relevant question interpretations resulting from the IQA pipeline.  
To facilitate efficient processing, we run the tools within each pipeline component in parallel. 

To select the tools for each pipeline component in the current realization of IQA, we conducted preliminary experiments. 

\subsubsection{Shallow Parser} 
\label{sec:shallow-parser}

We analyzed three independent shallow parsing tools, namely
MDP-Parser~\cite{Zafar2019MDPParser} developed in our previous work, SENNA~\cite{collobert2011natural} and a 
NLTK-based~\cite{bird2009natural} 
chunker implemented
using a classification-based sequential tagger.
MDP-Parser is a reinforcement learning-based approach to identify named entity and relation mentions in a distantly supervised setting.
In our preliminary experiments, we observed that MDP-Parser shows superior performance for shallow parsing compared to the other approaches \cite{Zafar2019MDPParser}. Furthermore, we did not observe any significant performance increase by adopting multiple tools at this pipeline step on the results of the entity and relation linking. 
Hence, we adopt MDP-Parser as the only tool in the Shallow Parser pipeline component. 

\subsection{Entity Linker}
\label{sec:entity-linker}

At this stage, we considered two state-of-the-art entity linking tools: TagMe~\cite{hasibi2016tagme} and EARL~\cite{dubey2018earl}. 
To further increase recall, we implemented an additional linking tool, which utilizes a character level n-gram representation of the information nuggets and performs linking between the information nuggets and the labels of entities in the knowledge graph using 3-gram similarity. 
We implemented this tool using an Apache Lucene index\footnote{https://lucene.apache.org}. 
In our preliminary experiments, we observed that entity linking results obtained by a combination of the 3-gram similarity and EARL subsume the results of TagMe. 
Thus we adopt the 3-gram similarity and EARL as two independent entity linking tools in the current realization of the IQA pipeline.

\subsubsection{Relation Linker}
\label{sec:relation-linker}

Relation linking is conducted analogously to the entity linking, using EARL and a word-matching similarity between the information nuggets
and the knowledge graph properties.

\subsubsection{Query Builder}
\label{sec:query-builder}

We adopt the SQG~\cite{zafar2018formal} tool developed in our previous work as the Query Builder component.

\subsection{Probability Estimation}
\label{sec:prob-estimation}

An estimation of the probability of a complete question interpretation $CQI=(QI, AT, QG)$ presented in Section \ref{sec:probabilityCQI}, requires estimation of the 
probabilities of the nugget interpretations $QI=\{ni_{1}, \ldots, ni_{r}\}$ and the query graph $QG$ of $CQI$. 

To estimate the probability of a nugget interpretation $P(ni_{i} |Q,\mathcal{KG})$, we adopt the confidence score of the pipeline component that generates this interpretation. 
We normalize the confidence scores using min-max scaling. 

The probability of the query graph
$P(QG|q_{NL},\mathcal{KG})$ is estimated using structural similarity of the query graph structure $QG$ and the user question $q_{NL}$.
To this extent, we use the Tree-LSTM based model of the SQG tool adopted as the query building component. SQG estimates the syntactical similarity of a candidate query that it generates to the parse tree structure of the input question $q_{NL}$. To estimate the probability of the query graph, we normalize the similarity score provided by SQG using the softmax function.

\subsection{IQA User Interface}
\label{sec:user-study}

\begin{figure*}[ht]
\includegraphics[width=\textwidth]{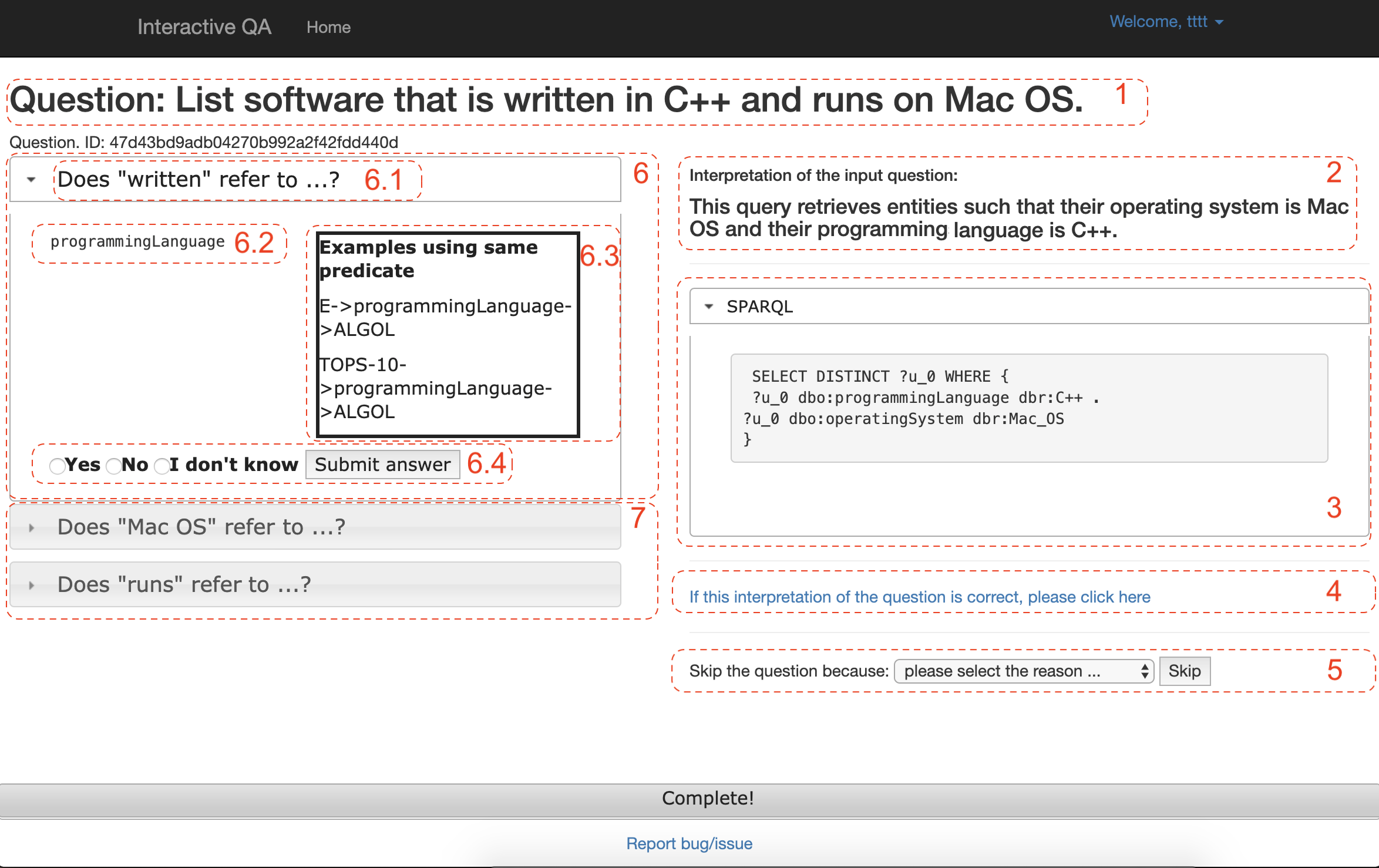}
\caption{User interface of IQA adopted in the user study.}
\label{fig:IQA_UI}
\end{figure*}

We implemented IQA prototype as a web application.
The user interface of IQA is exemplified in Figure \ref{fig:IQA_UI}. This interface is adopted in the user study described later in Section \ref{sec:ev-user}.
In general, IQA accepts any user-defined questions in a natural language. To enable a comparison of different approaches in the evaluation, during the user study, we adopted a controlled set of questions selected from the LC-QuAD dataset (we elaborate on the dataset generation later in Section~\ref{sec:ev-dataset}). In this section, we describe the user interface of IQA as it was presented to the users during the user study. 

First, the user signs up in the system. Then, the user logs in and starts the user study, where a page similar to Figure~\ref{fig:IQA_UI} is presented.
At the top of the interface, IQA displays the current question \textit{(\#1)}. 
On the right hand side, the top-ranked query is provided in its natural language representation \textit{(\#2)} (using SPARQL2NL~\cite{ngonga2013sparql2nl}) and a SPARQL representation \textit{(\#3)}. Using this part of the interface, the user can accept the top-ranked query  \textit{(\#4)}. Furthermore, if the user finds the presented question or the interaction options incomprehensible, the user can skip the question by choosing the corresponding reason and clicking on the skip button \textit{(\#5)}.

On the left-hand side, IQA provides the user with the current interaction option \textit{(\#6)}. 
The interaction option is expressed as an inquiry \textit{(\#6.1)} along with a candidate answer \textit{(\#6.2)}. The inquiry is in the form of ``Does \textit{'...'} refers to ...?'', where \textit{'...'} is a part of the original question. If applicable, a description and/or 
example of usages of the interaction option (in case the interaction option represents a relation) are displayed \textit{(\#6.3)}. The user can select from ``yes'', ``no'', and ``I don't know'' answers to accept or reject the interaction option displayed \textit{(\#6.4)}. The previously selected interaction options are listed below for user reference \textit{(\#7)}.

According to the user feedback, the interaction option and the top-ranked query are updated. The interaction continues until the user confirms the final semantic query or another termination criterion discussed in Section \ref{sec:interaction-process} is reached. 

To collect the usability feedback, IQA shows a dialog to the user upon completion of each question. In this dialog, IQA asks the user to rate the ease of use of the system. The usability rating is conducted on the scale from one to five, with one being difficult to use and five being easy to use. 
Finally, IQA presents the user with the next question.

A demo version of the IQA system is publicly available at \url{http://IQAdemo.sda.tech}.

%% file: 05a_evaluation_setup.tex
\section{Evaluation Setup}
\label{sec:evaluation-setup}

The goal of the evaluation is to demonstrate that IQA is competitive compared to both state-of-the-art interactive baselines and non-interactive approaches in terms of the effectiveness, efficiency, and usability for questions of different complexity. In this section, we describe the datasets and methods adopted for the evaluation.

\subsection{Knowledge Graph and Questions} 
\label{sec:ev-dataset}

We adopt LC-QuAD - an established dataset that contains $5,000$ complex questions for evaluation of SQA systems~\cite{LC-QuAD}\footnote{Available at \url{http://lc-quad.sda.tech/lcquad1.0.html}}. 
Overall, the LC-QuAD dataset contains questions in four complexity categories, i.e., questions that include 2-5 named entities and relations in the corresponding semantic queries. Consequently, we use the DBpedia dataset version 2016-10\footnote{Available at \url{https://wiki.dbpedia.org/downloads-2016-04}} as the underlying knowledge graph to be compatible with the semantic queries in the LC-QuAD dataset. 

To the best of our knowledge, Diefenbach et al.~\cite{DBLP:journals/corr/abs-1803-00832} provided the state-of-the-art results on the LC-QuAD dataset. 
Diefenbach et al. use a handcrafted vocabulary expansion for improving relation linking. This vocabulary is based on small parts of training data obtained from various Question Answering datasets, including SimpleQuestions and QALD-7. However, the authors did not clarify whether they use a portion of LC-QuAD to expand the vocabulary, as they do not provide any information regarding the train/test split for LC-QuAD. 
As the source code of \cite{DBLP:journals/corr/abs-1803-00832} is not available, we used the online API provided by the authors to reproduce their results within each complexity category. 
We noticed that 2,789 out 5,000 questions in LC-QuAD were not answerable due to the incompatibility of the DBpedia version used for the creation of LC-QuAD and the one used by the API. It was not possible to change the DBpedia version of the API; hence, to provide a fair comparison, we excluded the non-answerable questions and focused on the remaining 2,211 questions. On those questions, our computed $F_{1}$ score for WDAqua is 0.438, and their reported score is 0.46, which is similar.

For the oracle-based evaluation, we use the same subset of 2,211 LC-QuAD questions that we used for the evaluation of WDAqua.

\begin{figure}
  \centering
    \includegraphics[width=6cm]{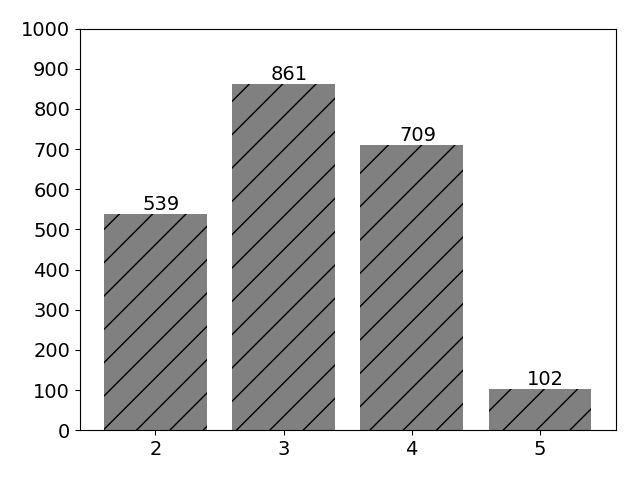}
    \caption{Question complexity distribution in the \textit{Oracle Test Questions} dataset. The X-Axis represents the complexity category. The Y-Axis represents the number of questions in the corresponding category. }
    \label{fig:comp_dist}
\end{figure}

We refer to this LC-QuAD subset as \textit{Oracle Test Questions}.
Figure \ref{fig:comp_dist} illustrates the 
distribution of the questions across the different complexity categories in the \textit{Oracle Test Questions} dataset.
As we can observe, the majority of the questions are in the complexity categories from two to four.

For the user evaluation, we select questions for which the IQA pipeline realized in this article can generate the semantic query specified in the LC-QuAD dataset (i.e., this query is generated by the IQA pipeline, but is not necessarily top-ranked). 
From this set, we randomly sample a set of questions, such that the number of questions in each complexity category is balanced. We refer to the set of 90 questions adopted in the user evaluation as \textit{User Test Questions}.

\subsection{Evaluation Metrics}
\label{sec:ev-metrics}

To assess the effectiveness, efficiency, and usability of the considered approaches, we adopt the metrics described in the following.

\subsubsection{Effectiveness}

To measure effectiveness, we choose 
\textit{Success Rate} and \textit{$F_{1}$ score}.

The \textit{Success Rate} is the percentage of the questions in a dataset for which the SQA approach can generate the intended semantic query. Note, that in case an approach generates several candidates, the intended semantic query does not have to be top-ranked.

The \textit{$F_{1}$ score} is the harmonic average of the precision and recall. 
Here, $F_{1}$ score corresponds to the 
\textit{Success Rate} at the top-1.

\subsubsection{Efficiency}

To measure efficiency, we adopt \textit{Interaction Cost}. 
We define the \textit{Interaction Cost} as the number of interaction options that the users need to consider before they can identify the semantic query that correctly interprets the question.
In the user evaluation, ``identify'' means that the user explicitly confirms the semantic query as correct.
In the oracle-based evaluation of interaction, ``identify'' means that the semantic query ranked at top-1 at the specific interaction round corresponds to the query given in the LC-QuAD dataset.  

In ranking-based approaches (e.g., in non-interactive baselines), 
the Interaction Cost is measured as the 
rank of the correct question interpretation, assuming 
that the user considers the semantic queries in their rank order. 

The lower values of the Interaction Cost correspond to the higher efficiency of an SQA system. 
The Interaction Cost of '1' corresponds to the case, where the intended semantic query is immediately shown (ranked at top-1) and confirmed by the user.

\subsubsection{Usability}

To assess usability, we design a rating scheme in which users can provide their feedback on the ease of use on the scale from one to five, with one being difficult to use and five being easy to use. 

\subsection{Evaluated Approaches}
\label{sec:ev-approaches}

In this work, we compare the performance of the SQA approaches and their configurations described in the following.

\subsubsection{IQA Configurations}
\label{sec:ev-iqa-config}

To assess the impact of the Option Gain proposed in this work as opposed to Information Gain, we compare two configurations of the proposed IQA approach: \textit{IQA-OG} and 
\textit{IQA-IG}.

In \textit{IQA-OG}, the interaction options are selected based on their Option Gain. 
We set the $\omega=1$ (see Equation \ref{eq:OptionGain}), such that both, Information Gain and usability of the options are taken into account equally.  

\textit{IQA-IG} is the interactive SQA method, where we take into account the Information Gain of the interaction options only. In this case, we set the parameter $\omega=0$ (see Equation \ref{eq:OptionGain}).

\subsubsection{Baselines}
\label{sec:ev-baselines}


To compare IQA to a state-of-the-art non-interactive SQA approach, we adopt NIB-WDAqua.

\textit{NIB-WDAqua: a Non-Interactive SQA Baseline using a state-of-the-art SQA approach.} 
In this case we take the state-of-the-art semantic SQA approach ``WDAqua-core1'' \cite{DBLP:journals/corr/abs-1803-00832} as a baseline. 
According to the recent evaluation on the Gerbil platform~\cite{usbeck2019benchmarking}\footnote{\url{http://gerbil-qa.aksw.org/gerbil/experiment?id=201805230002}}, an SQA benchmarking system,
``WDAqua-core1'' indicates the best performance concerning the LC-QuAD dataset adopted for the evaluation in this article. 
%
This baseline generates only one semantic query 
interpreting the user question. This query is provided by the authors of \cite{DBLP:journals/corr/abs-1803-00832} through their API\footnote{http://wdaqua-core1.univ-st-etienne.fr/gerbil}.  
%


To demonstrate the performance of the proposed IQA pipeline in 
the non-interactive settings, we use NIB-IQA.

\textit{NIB-IQA: a Non-Interactive SQA Baseline using the IQA pipeline.}
This baseline represents the IQA pipeline running without interaction. We assume that the IQA pipeline runs entirely automatically and outputs a ranked list of  
semantic queries at the end, where each semantic query interprets the user question in a specific way. 
To compute the Interaction Cost for the NIB-IQA baseline, we assume that the user considers the semantic queries generated by the pipeline in their rank order. In this case, the Interaction Cost corresponds to the rank of the semantic query in the resulting list.

%


To demonstrate the performance of the proposed interaction scheme compared to an interactive baseline, we consider SIB.

\textit{SIB: a Simple Interactive Baseline.}
This baseline involves user interaction after the execution of each SQA pipeline component. 
We assume that each pipeline component outputs a ranked list of interaction options (e.g., nugget interpretations). 
Furthermore, the Interaction Cost of each pipeline component is the rank of the first IO 
generated by this component that leads to the intended semantic query. 
This option is passed as an input to the next pipeline component. 
The overall Interaction Cost of the pipeline is the sum of the Interaction Cost over all the pipeline components.

\subsection{Evaluation Settings}
\label{sec:settings}

To assess the performance of IQA with respect to the evaluation metrics, facilitate comparison to the baselines and evaluate performance in the interaction involving human users, we performed 
an oracle-based evaluation and conducted a user study.

\subsubsection{Oracle-Based Evaluation}
\label{sec:ev-oracle}

To facilitate evaluation on an established large-scale dataset for Question Answering such as LC-QuAD, we adopt an oracle-based approach.

In particular, in the interaction process, we consider an interaction option to be \textit{correct} if the selection of this option can lead to 
the construction of the semantic query specified in the LC-QuAD dataset. 
In the automatic evaluation, we simulate the user interaction process by 
letting the system automatically accept the first correct option suggested by the adopted SQA method.  
This corresponds to the assumption that the user would 
always select the correct option if this option is suggested by the system.

\subsubsection{User Study}
\label{sec:ev-user}

To better understand the impact of the proposed Option Gain metric on the effectiveness, efficiency, and usability of the IQA scheme (IQA-OG) in comparison to the interaction based on Information Gain (IQA-IG) when involving human users, we conducted a user study.

To enable evaluation of the proposed approach in the controlled settings, we adopted a homogeneous user group with  
15 post-graduate computer science students. We envision evaluation with other user groups to be an important part of the future research. 

At the beginning of the study, the authors of the article have briefly introduced the users to the IQA system. 
During the study, each user evaluated 12 questions on average (3 questions in 4 complexity categories). 
On average, users spent 30 minutes to conduct the study. 
For the configuration of the user study, the following rules were applied:
\begin{itemize}
    \item To facilitate a comparison of the methods, each question is evaluated using two IQA configurations: IQA-OG and IQA-IG.
    \item During the study, each user interacts with the system using one fixed interaction configuration, either IQA-OG or IQA-IG.
    \item The user does not receive the same question twice.
    \item The user can mark a question as incomprehensible. The question marked by any user is removed from the \textit{User Test Questions} set.

    The remaining set of \textit{User Test Questions} contains 80 questions.
\end{itemize}

Figure \ref{fig:IQA_UI} illustrates the user interface of IQA adopted in the user study
with an example question from the \textit{User Test Questions} set.

User study results are discussed in Section \ref{sec:user_evaluation_results}.

\subsection{Reproducibility} 
\label{sec:ev-repro}

To support the reproducibility of results and facilitate further research, 
we make the software and the data adopted in the evaluation available as follows.
The source code of the interactive query construction is available on our GitHub repository\footnote{\label{git_iqa}\url{https://github.com/AskNowQA/InteractiveQA}}. 
Similarly, the source code of the MDP-Parser\footnote{\url{https://github.com/AskNowQA/DeepShallowParsingQA}}, SQG\footnote{\url{https://github.com/AskNowQA/SQG}} as well EARL\footnote{\url{https://github.com/AskNowQA/EARL}} are available on GitHub.
Furthermore, the experimental results for the oracle evaluation are provided at our GitHub repository~\textsuperscript{\ref{git_iqa}}.

%% file: 05b_evaluation_results.tex
\section{Evaluation Results}
\label{sec:evaluation_results}

In this section, we present the results of the oracle-based evaluation and the user study.

\subsection{Oracle-based Evaluation Results}
\label{sec:automatic_,evaluation_results}

We assess the effectiveness and efficiency of the proposed IQA approach on an established large-scale LC-QuAD dataset using the oracle-based evaluation.

\subsubsection{Effectiveness Results in the Oracle-based Evaluation}

\begin{figure}
  \centering
    \includegraphics[width=0.7\columnwidth]{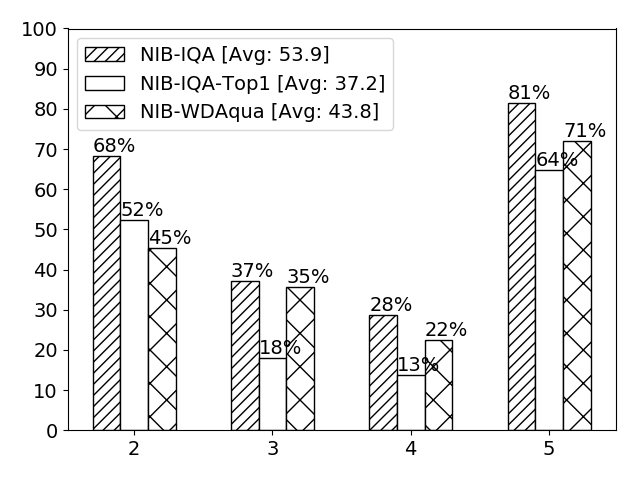}
    \caption{Success Rate of the non-interactive baselines NIB-IQA, NIB-IQA-Top-1 and NIB-WDAqua for the questions in the \textit{Oracle Test Questions} dataset. The X-Axis represents the complexity category. The Y-Axis represent the Success Rate. 
    }
    \label{fig:success-rate}
\end{figure}

Figure \ref{fig:success-rate} presents the Success Rate 
of the non-interactive baselines. 
The NIB-WDAqua baseline that represents a state-of-the-art Semantic Question Answering approach \cite{DBLP:journals/corr/abs-1803-00832} generates only the top-1 semantic query. 
The NIB-IQA baseline, i.e., a non-interactive version of the proposed IQA approach, generates multiple candidate semantic queries. 
With NIB-IQA-Top-1, we consider only the top-1 query generated by the NIB-IQA baseline. 

As we can observe in Figure \ref{fig:success-rate}, 
NIB-IQA outperforms the NIB-WDAqua baseline in terms of Success Rate
in all complexity categories. 
Whereas the NIB-WDAqua outperforms the NIB-IQA with respect to the top-1 query (i.e., the NIB-IQA-Top1 baseline), the overall Success Rate of NIB-IQA is higher than the Success Rate of the NIB-WDAqua.
This is because NIB-IQA generates multiple relevant question interpretations, whereas NIB-WDAqua does not provide such functionality
and returns only one top-ranked query.

As expected, we can observe that the overall performance of all non-interactive question answering pipelines degrades with the increasing complexity of the questions in the categories 2-4. 
A special case is the Success Rate in the complexity category 5, where the questions follow a similar pattern, which makes it relatively easy for all considered SQA systems to construct the corresponding semantic query.  

As we can observe in Figure~\ref{fig:success-rate}, in the complexity category 2, 68\% of the queries are answerable by NIB-IQA (i.e., the intended query is constructed by the IQA pipeline), whereas this query is ranked as top-1 (NIB-IQA-Top1) only in 52\% of the cases. 
Overall, the difference between the NIB-IQA and NIB-IQA-Top1 is 16.7 percentage points on average across the complexity categories. 

The approach proposed in this article fills this gap, such that the difference between NIB-IQA and NIB-IQA-Top1 is reduced through interaction 
%
(as will be demonstrated later in the results of the oracle-based evaluation in 
Section \ref{sec:oracle-based-evaluation} and the discussion of the user study presented in Section \ref{sec:user_evaluation_results}).
I.e., with interaction, the Success Rate of the NIB-IQA-Top1 will increase and can reach the Success Rate of NIB-IQA, 
outperforming the NIB-WDAqua baseline.

\begin{figure*}
    \centering
    \begin{subfigure}[b]{0.45\textwidth}
        \centering
        \includegraphics[width=5.5cm]
        {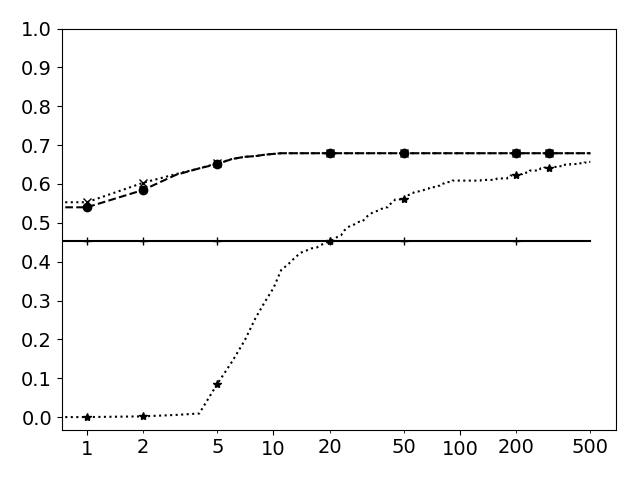}
        \caption{ $F_{1}$ score for the questions with complexity 2}
        \label{fig:Complexity}
    \end{subfigure}
    \begin{subfigure}[b]{0.45\textwidth}
        \centering
        \includegraphics[width=5.5cm]
        {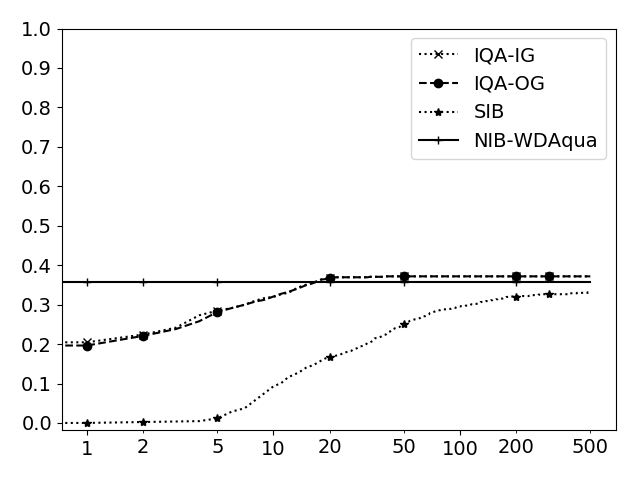}
        \caption{$F_{1}$ score for the questions with complexity 3}
        \label{fig:inter_inc_f1_3}
    \end{subfigure}
    \hspace{8pt}%
    \begin{subfigure}[b]{0.45\textwidth}
        \centering
        \includegraphics[width=5.5cm]
        {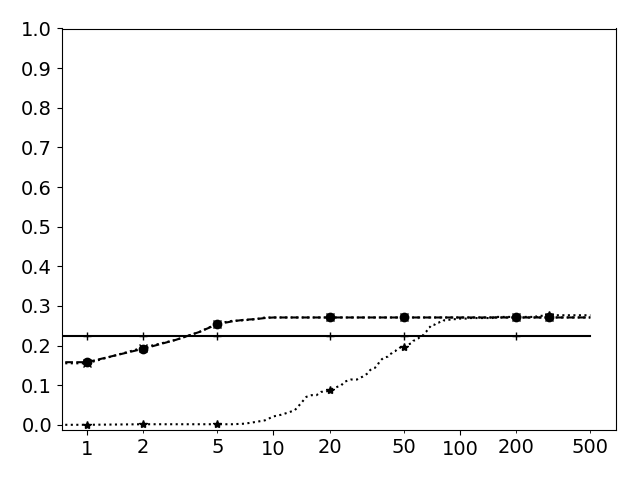}
        \caption{$F_{1}$ score for the questions with complexity 4}
        \label{fig:inter_inc_f1_4}
    \end{subfigure}
    \begin{subfigure}[b]{0.45\textwidth}
        \centering
        \includegraphics[width=5.5cm]
        {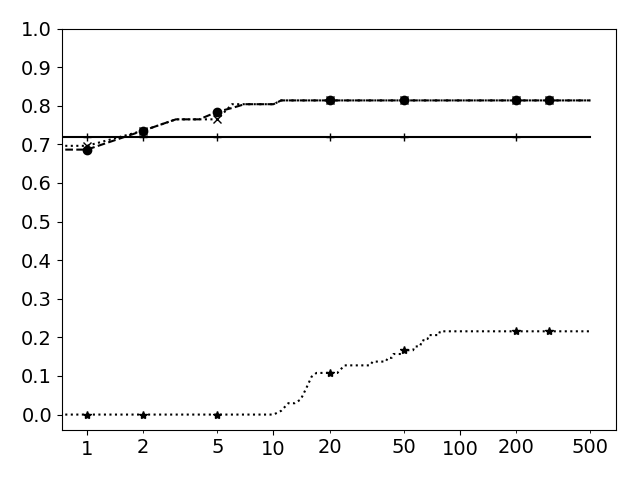}
        \caption{$F_{1}$ score for the questions with complexity 5}
        \label{fig:inter_inc_f1_5}
    \end{subfigure}
    \caption{Increase in $F_{1}$ score during the interaction process in the oracle-based evaluation. The X-Axis represents the number of interactions on a log scale. The Y-Axis represents the $F_{1}$ score.
    }
    \label{fig:inter_inc_f1}
\end{figure*}

Figure \ref{fig:inter_inc_f1} shows the $F_{1}$ score obtained using different methods and the evolution of the $F_{1}$ score during the interaction process achieved due to the reduction of the question interpretation space.
The X-Axis represents the number of interactions on a log scale. 
The Y-Axis represents the $F_{1}$ score.
We show the results for the questions of different complexity in separate sub-figures of Figure \ref{fig:inter_inc_f1}.
%

The baseline method NIB-WDAqua conducts only one interaction with the user, i.e., it generates the top-1 semantic query that interprets the question \cite{DBLP:journals/corr/abs-1803-00832}. This semantic query remains unchanged in the interaction process (the API of the \cite{DBLP:journals/corr/abs-1803-00832} does not provide any other interpretations); therefore, the result of the NIB-WDAqua baseline is represented as a straight line in Figure \ref{fig:inter_inc_f1}.

%
As expected, given the results presented above, the NIB-WDAqua baseline shows the best results at the very beginning of the interaction process in categories 3-5. However, after a few interactions, the NIB-WDAqua baseline is outperformed by other approaches in all complexity categories.

The interactive configurations IQA-OG and IQA-IG of the proposed approach, demonstrate similar performance. 

The SIB interactive baseline shows the worst performance across the approaches presented in Figure \ref{fig:inter_inc_f1} in all complexity categories. SIB implements an extensive interaction strategy and requests user feedback at every pipeline step. This result confirms our intuition that interaction alone is not sufficient to 
construct the intended question interpretation efficiently.
The significant differences between SIB and the informed interaction strategy of IQA (reflected by IQA-OG and IQA-IG) highlight the clear advantage of our proposed approach in comparison to this baseline.

\subsubsection{Efficiency Results in the Oracle-based Evaluation}
\label{sec:oracle-based-evaluation}

\begin{figure}
  \centering
    \includegraphics[width=0.7\columnwidth]{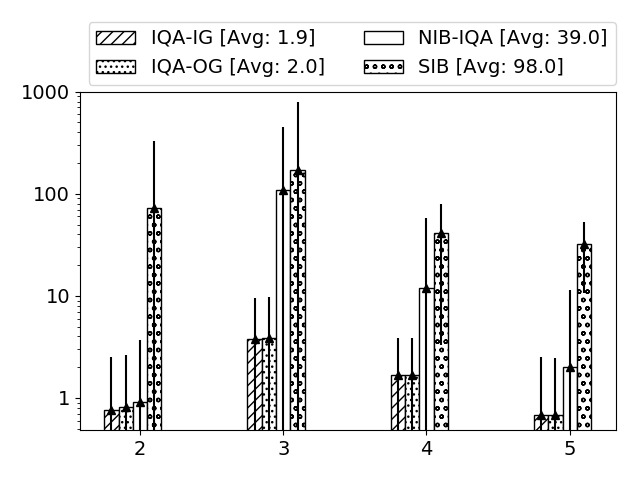}
    \caption{Interaction Cost and std. deviation of different approaches in the oracle-based evaluation. The X-Axis represents the complexity category of the question. The Y-Axis represents the Interaction Cost. The Y-Axis is logarithmic.
   The bars represent the results of the proposed interactive approaches IQA-IG and IQA-OG as well as of the baselines NIB-IQA and SIB.
   }
    \label{fig:inter-cost-automatic}
\end{figure}

Figure \ref{fig:inter-cost-automatic} presents the 
Interaction Cost and the standard deviation of the considered approaches achieved in the different complexity categories in the oracle-based evaluation over the \textit{Oracle Test Questions} dataset. 

As we can observe in Figure \ref{fig:inter-cost-automatic}, IQA-IG, and IQA-OG have significantly lower Interaction Cost compared to the NIB-IQA and SIB baselines. The Interaction Cost of IQA-OG and IQA-IG in the oracle-based settings are equivalent.
This result demonstrates that an interactive approach based on Option Gain or Information Gain can significantly reduce the Interaction Cost compared to the baselines.
This result also illustrates that although multiple outputs as produced by the NIB-IQA baseline can facilitate interaction, if taken without further optimization, such multiple outputs are not sufficient to effectively reduce the Interaction Cost.

\subsection{User Study Results}
\label{sec:user_evaluation_results}

The goal of the user study is to assess the performance of IQA-OG and IQA-IG approaches in terms of their efficiency, usability, and effectiveness
in the interaction involving human users. 
In this section, we present the results of the user study.

\subsubsection{Efficiency}
\begin{figure*}
    \begin{subfigure}[b]{0.45\textwidth}
        \centering
        \includegraphics[width=5.5cm]{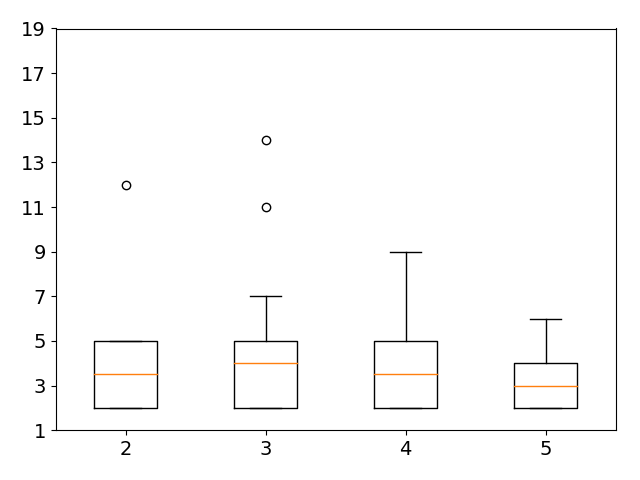}
        \caption{IQA-IG}
        \label{fig:ig}
    \end{subfigure}
    \begin{subfigure}[b]{0.45\textwidth}
        \centering
        \includegraphics[width=5.5cm]{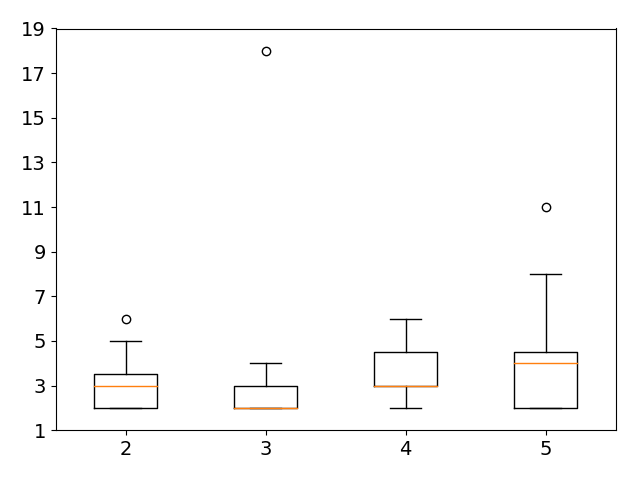}
        \caption{IQA-OG}
        \label{fig:og}
    \end{subfigure}
    \caption{Interaction Cost of IQA-IG and IQA-OG in the user study in a boxplot representation.
        The X-Axis represents the complexity category.
        The Y-Axis represents the Interaction Cost.    }
    \label{fig:ig-og}
\end{figure*}

We measure the efficiency of interaction using Interaction Cost. Figure \ref{fig:ig-og} presents the Interaction Cost observed in the user evaluation for the questions of different complexity while using IQA-OG and IQA-IG configurations of the proposed approach.

Overall, the Interaction Cost of both IQA-OG and IQA-IG is relatively low, with 3.8 interactions on average for IQA-IG and 3.6 for IQA-OG. As we can observe in Figure \ref{fig:ig-og}, both approaches indicate slight variations. However, the results of the paired t-test show that these differences are not statistically significant. We conclude that both methods, IQA-OG and IQA-IG, are equivalent in terms of efficiency. 

Compared to the results of the oracle-based evaluation, the Interaction Cost observed in the user study is slightly higher. The average Interaction Cost in the oracle-based evaluation presented in Figure \ref{fig:inter-cost-automatic} is 1.9-2.0, whereas, in the user study, we observed 3.6-3.8 interactions on average.  
This is because, in comparison to the oracle-based setting, the users do not always immediately confirm the top-ranked query once it is shown, but may continue the interaction process.

\begin{figure*}[ht]
    \begin{subfigure}[b]{0.45\textwidth}
        \centering
        \includegraphics[width=5.5cm]{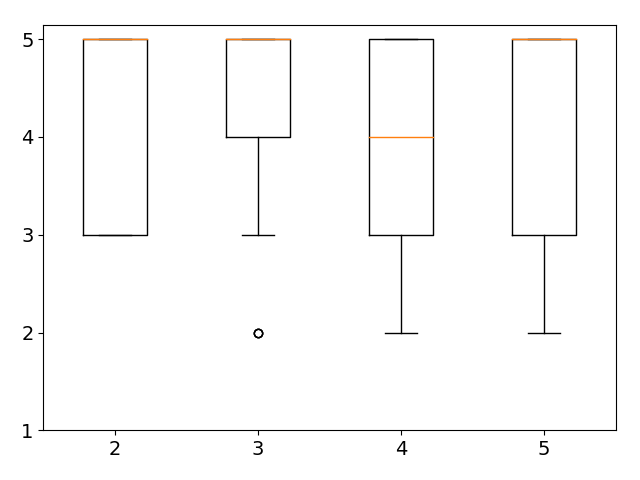}
        \caption{IQA-IG}
        \label{fig:fig:user_feedback_ig}
    \end{subfigure}
    \begin{subfigure}[b]{0.45\textwidth}
        \centering
        \includegraphics[width=5.5cm]{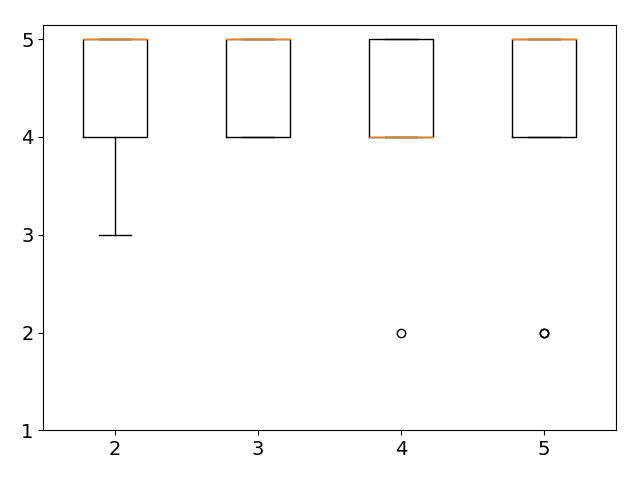}
        \caption{IQA-OG}
        \label{fig:user_feedback_og}
    \end{subfigure}
    \caption{User rating on IQA usability in a boxplot representation. Average rating of IQA-IG=4.13; average rating of IQA-OG=4.40.}
    \label{fig:user_feedback}
\end{figure*}

\subsubsection{Usability}

Figure \ref{fig:user_feedback} presents the usability results of IQA-IG and IQA-OG computed using user ratings. 
The average user rating is 4.13 for IQA-IG and 4.40 for IQA-OG. According to the paired t-test, this result is statistically significant ($p<.05$).
As we can observe, the scores obtained by IQA-IG are not only lower on average, but also indicate much higher variation. 
We conclude that IQA-OG outperforms IQA-IG with respect to the ease of use.

\subsubsection{Effectiveness}
\label{sec:user-benchmark-differences}

We assess the effectiveness of the interaction scheme in the user evaluation as the accuracy in the construction of the intended semantic queries.

As discussed in Section \ref{sec:ev-user}, to complete the interaction process for each question, the user had to explicitly confirm if the constructed query correctly reflected the intention of the question. 
The query confirmed by the user can be different from the semantic query specified in the LC-QuAD dataset.  
In this section, we discuss the observed deviations between the queries confirmed by the users and the queries specified in the LC-QuAD dataset.

\begin{figure*}
    \begin{subfigure}[b]{0.3\textwidth}
        \centering
        \includegraphics[width=5.5cm]{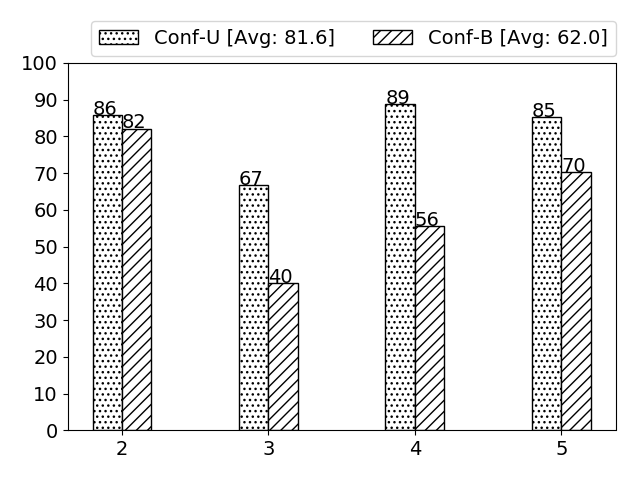}
        \caption{IQA-IG}
        \label{fig:user_vs_benchmark-ig}
    \end{subfigure}
    \begin{subfigure}[b]{0.3\textwidth}
        \centering
        \includegraphics[width=5.5cm]{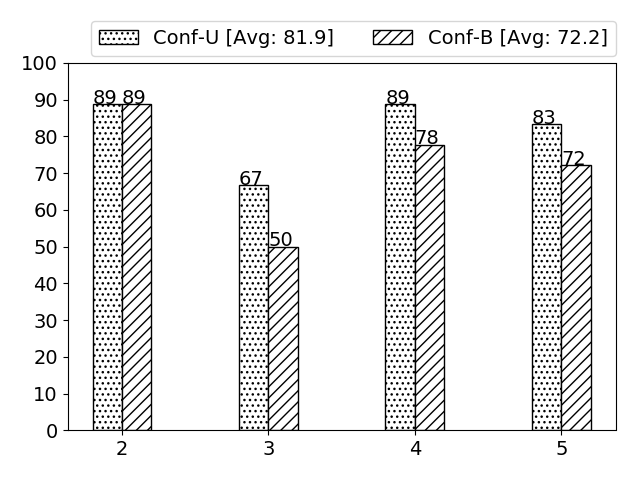}
        \caption{IQA-OG}
        \label{fig:user_vs_benchmark-og}
    \end{subfigure}
    \begin{subfigure}[b]{0.3\textwidth}
        \centering
        \includegraphics[width=5.5cm]{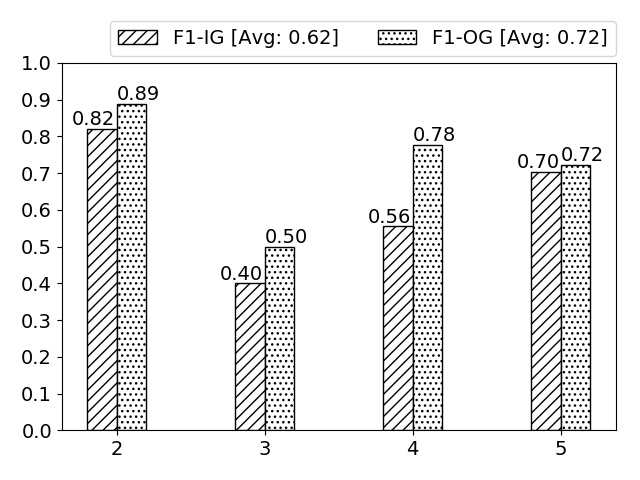}
        \caption{$F_{1}$ score of IQA-IG and IQA-OG}
        \label{fig:f1-igog}
    \end{subfigure}
    \caption{Accuracy of the user judgments vs. the LC-QuAD dataset. The X-Axis represents query complexity. In \ref{fig:user_vs_benchmark-ig} and \ref{fig:user_vs_benchmark-og}, the Y-Axis represents the ratio of questions for which the semantic query was confirmed by the user (Conf-U) and the ratio of queries, which are equivalent to the LC-QuAD dataset (Conf-B) obtained using IQA-IG and IQA-OG.
    In \ref{fig:f1-igog}, the Y-Axis represents the $F_{1}$ score achieved by the users
    using the IQA-IG and IQA-OG configurations.
    }
    \label{fig:user-vs-benchmark-diff}
\end{figure*}

Figures \ref{fig:user_vs_benchmark-ig} and \ref{fig:user_vs_benchmark-og} present the ratio of questions in different complexity categories that are: 
1) confirmed by the users as correct (Conf-U), and 
2) confirmed by the users as correct and also exactly correspond to the semantic query 
in the LC-QuAD dataset (Conf-B).
We present these statistics for the IQA-OG and IQA-IG configurations.

As we can observe in Figures \ref{fig:user_vs_benchmark-ig} and \ref{fig:user_vs_benchmark-og}, the users have confirmed semantic  
queries that were not contained 
in the LC-QuAD dataset in all complexity categories, whereas the differences between Conf-U and Conf-B are much smaller for IQA-OG.
Note that Conf-B directly corresponds to the $F_{1}$ score presented in Figure \ref{fig:f1-igog}. 

Figure~\ref{fig:f1-igog} indicates that the queries constructed using IQA-OG are more accurate, which is likely due to the interaction options adopted by this approach that can be better understandable by users. The average percentage of queries constructed by the users and confirmed by the LC-QuAD dataset is 62.0\% for IQA-IG and 72.2\% for IQA-OG. We observe that IQA-OG consistently outperforms IQA-IG in all complexity categories, with an average improvement of 10 percentage points in $F_{1}$ score. 

This observation again indicates that IQA-OG that takes usability of the options into account can facilitate more effective user interaction than an interaction approach based solely on the Information Gain. 

Overall, compared to IQA-IG, IQA-OG leads to more intuitive user interaction that facilitates the user to answer the questions more effectively, within the same number of interactions.

\begin{figure}
  \centering
    \includegraphics[width=0.7\columnwidth]{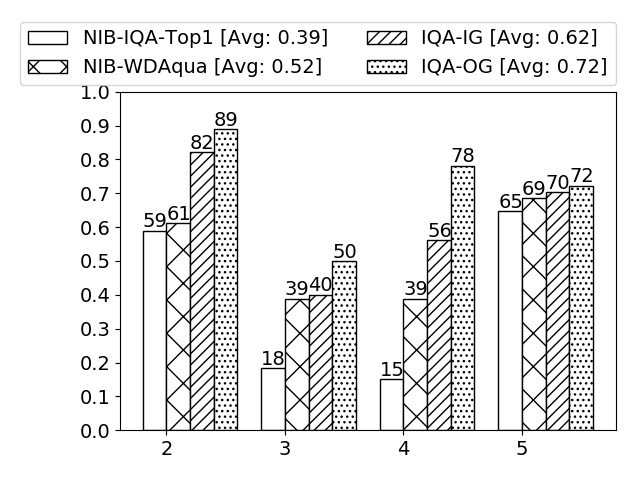}
    \caption{The X-Axis represents query complexity. Y-Axis represents the $F_{1}$ score achieved by different approaches on the \textit{User Test Questions}. 
    IQA-IG and IQA-OG correspond to the user study results. 
   }
    \label{fig:f1-IGOG-top1-wd}
\end{figure}

Figure~\ref{fig:f1-IGOG-top1-wd} depicts the $F_{1}$ scores achieved on the \textit{User Test Questions} by different approaches.
IQA-IG and IQA-OG scores correspond to the user study results. 
NIB-WDAqua and NIB-IQA-Top1 are the baseline results achieved on the same dataset. 
As we can observe, the proposed interactive approach 
outperforms the best performing non-interactive baseline NIB-WDAqua 
concerning the $F_{1}$ scores in all complexity categories.
The average $F_{1}$ score of IQA-IG is 0.62, which is an increase of 10 percentage points compared to the NIB-WDAqua baseline that obtains $F_{1}=0.52$ on average on this dataset. With the IQA-OG, we achieve an $F_{1}=0.72$, which is 20 percentage points higher than the $F_{1}$ score of the NIB-WDAqua baseline.

\subsubsection{Error Analysis}
As for the failed questions, on average, 11\% were rejected by the users due to incomprehensible questions or interaction options,
whereas 15\% failed as the users did not confirm the semantic query resulting from the interaction process.

To better understand the differences between the queries constructed and accepted by the users 
and the semantic queries in the LC-QuAD dataset, we conducted a manual inspection of all results where such deviation occurred. 
Overall, we observed several reasons for deviations, including: 
\begin{itemize}
\item [R1] The LC-QuAD interpretation is too restrictive: There exist several possible semantic interpretations for a question, 
and LC-QuAD only includes one such interpretation. For example, this can be observed in the case of synonymous relations, 
or inclusion/omission of the \textit{rdf:type} statements in the semantic query that do not affect the results.
\item [R2] The user makes a mistake or fails to understand the specific differences between the intended interpretation and 
the interpretation suggested by the system. For example, this can happen in case of similar entities, or a wrong interpretation of the relation direction by the user.
\item [R3] The user selects a different answer type. For example, the user can accept a SELECT query instead of an ASK query specified in LC-QuAD.
\end{itemize}
We provide an overview of the typical differences, their frequency and the corresponding examples in 
Table \ref{tab:differences}. As we can observe, the most frequent reasons for the deviations are the synonymous relations (R1, in 43.4\%), wrong relations (R2, in 19.5\%), and the differences in the answer types (R3, in 19.5\%).

\begin{table*}[t]
\centering
\caption{Differences of the user interpretation and LC-QuAD.}
\label{tab:differences}
\begin{adjustbox}{width=\textwidth}

\begin{tabular}{@{}llrl@{}}
\toprule
\textbf{Reason} & \textbf{Differences}                & \textbf{
\%} & \textbf{Example} \\ \midrule
R1 & Synonymous relations  &  43.4  &
\small{Q: Name the home stadium of FC Spartak Moscow?} \\ &&& dbp:stadium vs. dbo:homeStadium   \\
R1 & Completeness & 8.6  & \small{Q: Miguel de Cervantes wrote the musical extended from which book?} \\ &of the semantic query && \small{SELECT ?u WHERE \{ ?u dbo:author dbr:Miguel\_de\_Cervantes \}}
\\
&&&
\small{SELECT ?u WHERE \{ ?u dbo:author dbr:Miguel\_de\_Cervantes.}  
\\
&&& \hspace{3.8cm}\small{?u rdf:type dbo:Book \}}
\\
R2 & Similar entities  &  4.5  & \small{Q: In which state is Red Willow Creek?} \\ &&& dbr:Willow\_Creek\_mine vs. dbr:Red\_Willow\_Creek \\
R2 & Wrong relation & 19.5 & \small{Q: List the producer of the TV shows whose company is HBO.} \\
&&& dbo:distributor vs. dbo:company\\
R2 & Structural differences & 4.5  & 
\small{Q: Who are the predecessors of John Randolph of Roanoke?} \\&in the semantic query &&
\small{SELECT ?u WHERE \{ dbr:John\_Randolph\_of\_Roanoke dbp:predecessor ?u\}} \\ &&&
\small{SELECT ?u WHERE \{  ?u dbp:predecessor dbr:John\_Randolph\_of\_Roanoke\}}
\\
R3 & Differences in the  & 19.5 & SELECT ?u WHERE ... vs. ASK WHERE ... \\ 
 & answer type  &   &   \\ 
\bottomrule
\end{tabular}%
\end{adjustbox}
\end{table*}

\subsubsection{User Feedback}
\label{sec:user-feedback}

After the evaluation session, we requested the users to provide unstructured feedback regarding any issues they observed or comments they had. 

Overall, the users reported a positive experience with the IQA system. The typical issues reported by the users included sometimes unclear formulation of the questions in the LC-QuAD dataset, understandability of interaction options in some categories, and of natural language formulation of complex SPARQL queries. 

As reported by the users, the LC-QuAD dataset contains some questions with linguistic issues. In cases where these issues affected the understandability of questions, the users could skip the question, as mentioned above. We consider such questions as failed in our results.

The users also reported occasional difficulties in understanding the semantics of some of the interaction options, in particular concerning the options representing relations and question types. This observation confirms our assumption used as a basis for the Option Gain computation that the usability of different interaction option types varies. 

Finally, the users reported that some of the natural language representations of the SPARQL queries, especially in the context of the more complex questions, were difficult to understand. The generation of the natural language representations for the user interface is not in the scope of this work; in the IQA prototype implementation, we generated such representations using state-of-the-art tools. However, this observation indicates the need for future work in this area.

%% file: 06_relatedwork.tex
\section{Related Work}
\label{sec:relatedwork}

Interactive methods to obtain user feedback have been adopted in Semantic Question Answering systems as well as in keyword search and natural language interfaces for structured data. In this section, we briefly summarize the differences between IQA and these approaches.

\subsection{Interactive Keyword Search over Relational Data}

In our previous work we proposed FreeQ - an interactive keyword search approach for relational databases \cite{Demidova:2013},\cite{Demidova:2012},\cite{Demidova:FreeQ}. 
FreeQ generates interaction options using a relational database schema and a mapping between the schema and an external ontology (utilizing, e.g., YAGO+F \cite{Demidova:YAGO+F}). User interaction in FreeQ is based on Information Gain.
Whereas IQA builds upon our previous work in the area of interactive keyword search, in this article, we target a more complex problem of Semantic Question Answering. The input questions are more complex than keyword queries supported by FreeQ, so are the corresponding SQA pipelines. 
IQA addresses these challenges through a novel interaction scheme dedicated to Semantic Question Answering. In particular, in IQA, we developed an interaction scheme for generic Semantic Question Answering pipelines. Furthermore, we introduced the notion of Option Gain that takes the usability of interaction options into account. As our evaluation demonstrates, these contributions lead to significant improvements in terms of usability and effectiveness, while maintaining low interaction cost.

\subsection{Semantic Question Answering}

Semantic Question Answering over knowledge graphs is a difficult problem \cite{hoffner2017survey, diefenbach2017coresurvey}.
Although SQA systems over simple questions have improved in recent years \cite{lukovnikov2017neural, bordes2015large}, solving complex questions \cite{dubey2016asknow, stagg, LC-QuAD} remains a difficult task. 
For example, the ``WDAqua-core1'' system \cite{DBLP:journals/corr/abs-1803-00832}, currently the best performing over the LC-QuAD dataset containing complex queries, only achieves $F_{1}=0.46$. 
SQA systems usually suffer a performance loss due to the wrong interpretations during the entity linking \cite{dubey2018earl, hasibi2016tagme}, relation linking, and query building \cite{zafar2018formal} stages.
These systems are typically optimized to produce one intended interpretation. 
In contrast to IQA, such systems do not support user feedback to refine their results.

\subsection{Interactive Question Answering Systems}
Existing SQA and search systems over knowledge graphs employ user feedback and additional input 
to improve disambiguation of the questions directly, or to generate training data. 
For example, Exemplar Queries~\cite{Mottin:2014} employs a user query as an example to search for similar structures. Su et al. \cite{Su:2015} exploit relevance feedback to tune ranking functions in knowledge graph search. GQBE \cite{Jayaram:2015} takes a question and an example relation as input and searches for similar graph patterns. Zheng et al. \cite{Zheng:2017} conduct interactive graph search and let users verify the ambiguities in entity linking, relation linking, and query building. IMPROVE-QA \cite{Zhang:2018} asks the users to correct the output of the training question to improve relation linking and query building process by learning from user interaction. 
In contrast to existing SQA systems that adopt interaction, IQA explicitly addresses the usability aspects of user interaction through Option Gain and utilizes a broader range of interaction options. 

\subsection{Other Interactive Approaches using Knowledge Graphs}

Sparklis \cite{Ferre17} is an exploration-based approach that allows users to build SPARQL queries interactively. In contrast, IQA is a Semantic Question Answering approach that adopts interaction for the disambiguation of user questions.
EventKG+TL facilitates interactive generation of multilingual event timelines from a knowledge graph \cite{GottschalkD18a}.
Conversational approaches such as CuriousCat \cite{BradeskoWSHGM17} provide another type of interaction. 
These approaches address other objectives, including, for example, knowledge acquisition in a dialog.

\subsection{Interactive Semantic Parsing}

Several works on interactive semantic parsing adopt user feedback as a training signal to resolve utterance ambiguity and enhance parsing accuracy. These approaches translate the natural language to formal domain-specific representations, including database queries \cite{Li:2014}, API calls \cite{SuAWW18}, and If-Then programs \cite{yao2019interactive}. 
Semantic parsing approaches that target translation of natural language into SQL queries for relational databases, such as, for example, \cite{Li:2014}, \cite{YavuzGSY18}, \cite{yao2019model} are the most related to our work. 
Approaches in this area are typically limited to rather small database schemas or simple query patterns. For example, \cite{Li:2014} performs evaluation on the Microsoft Academic Search (MAS) dataset that includes only eight relations. DialSQL \cite{YavuzGSY18} and MISP \cite{yao2019model} adopt the WikiSQL dataset that contains rather simple queries. 
In contrast, interactive SQA systems such as IQA aim to generate semantic queries for knowledge graphs that are much larger in scale, including thousands of concepts and relations, while enabling complex queries. This large scale poses additional challenges concerning the scalability and the interaction cost. 
Furthermore, approaches to interactive semantic parsing in databases invoke interaction based on ambiguity \cite{Li:2014} or error detection \cite{yao2019model}, and do not address usability aspects.

%% file: 07_conclusion.tex
\section{Conclusion}
\label{sec:conclusion}

In this article, we presented IQA - a novel interactive approach to Semantic Question Answering. 
We formalized the concept of a Semantic Question Answering pipeline and proposed a novel probabilistic user interaction scheme. This scheme aims to facilitate the user to effectively identify the intended semantic query while increasing the usability of interaction and minimizing the interaction cost. 
Interaction options utilized by the IQA belong to several categories, including interpretations of entities and relations, superclasses and types of entities, answer types, and semantic queries. In the interaction process, these options are determined based on their Option Gain, which takes into account the usability and efficiency of the options. 

To evaluate the effectiveness, efficiency, and usability of the proposed user interaction scheme, we conducted an extensive oracle-based experimental evaluation and a user study. Our experimental results over LC-QUAD, an established dataset in the assessment of SQA systems, demonstrate that IQA can significantly increase the effectiveness of SQA for complex questions while maintaining high usability of interaction and incurring only a small interaction cost.

We observed that an interaction strategy IQA-OG based on the Option Gain leads to higher user satisfaction compared to IQA-IG optimized for efficiency only. Furthermore, IQA-OG leads to the higher effectiveness of the user interaction, as reflected by the higher ratio of successfully constructed semantic queries. This improvement is reflected in the $F_1$ score that outperforms the interaction strategy based on the Information Gain by ten percentage points. Compared to the non-interactive baselines, IQA-OG achieves up to 20 percentage points improvement on the subset of LC-QUAD utilized in the user evaluation. 

We believe that this improvement is due to the less complex and thus better understandable interaction options adopted by the IQA-OG, which help to reduce potential errors. 

In principle, the IQA interaction scheme is applicable on top of any Semantic Question Answering pipeline that realizes the generic architecture formalized in this article. In particular, we support variations of SQA pipelines in the linking step, such that there can be a single joint linking step for entities and relations or multiple individual linking steps. This way, the IQA interaction approach can be applied to a broader range of existing SQA frameworks. 

In our future work, we plan to further develop the proposed approach to better support user interaction in multilingual settings.